\documentclass[pre,aps,twocolumn]{revtex4-2}
\usepackage{amsfonts}
\usepackage{times}
\usepackage{amsmath, amsthm}
\usepackage[colorlinks=true,citecolor=blue]{hyperref}
\usepackage[final]{graphicx}
\usepackage{amssymb, algorithm, algpseudocode}
\algnewcommand\algorithmicforeach{\textbf{for each}}
\algdef{S}[FOR]{ForEach}[1]{\algorithmicforeach\ #1\ \algorithmicdo}
\usepackage[title]{appendix}
\usepackage{xcolor,comment}

\newcommand{\sech}{\mathrm{sech}}

\begin{document}

\title{Nonlinear dispersive waves in the discrete modified KdV equation}

\author{S. Yang}
\affiliation{Department of Mathematics and Statistics, University
of Massachusetts, Amherst, Massachusetts 01003-4515, USA}

\date{\today}

\begin{abstract}
    In this paper, we study the nonlinear dispersive waves including the rarefaction and dispersive shock waves in the discrete modified KdV equation through the numerical simulations of the dispersive Riemann problems. In particular, we propose distinct quasi-continuum models to approximate both the spatial profiles and distinct edge features of these two specific dispersive wave structures. Whitham analysis is performed to construct a closed system of partial differential equations which describe the slowly-varying dynamics of all the relevant parameters associated with the periodic traveling waves of the proposed quasi-continuum models. We then perform reduction on such modulation system to obtain a system of two simple-wave ordinary differential equations which lead to the DSW-fitting method that shall provide useful theoretical insights on different edge characteristics of the dispersive shock waves. Furthermore, we compute analytically the self-similar solutions corresponding to the dispersionless systems of the quasi-continuum models, which can be utilized to approximate the numerically observed rarefaction waves of the discrete mKdV equation. A systematic numerical comparison of these theoretical findings with their associated numerical counterparts finally demonstrate the good performance of the proposed quasi-continuum models in approximating both nonlinear dispersive wave patterns.
\end{abstract}

\maketitle

\section{Introduction}

Dispersive shock wave (DSW) \cite{Hoefer:2009}, which is a non-stationary dispersive wave structure, can be numerically observed in the simulations of a variety of mathematical physics models including the famous nonlinear Schr\"odinger equation \cite{yang2026dispersiveshockwavesperiodic,mohapatra2026dambreaksdiscretenonlinear,PhysRevA.110.023304}, the granular crystal lattice \cite{Yang_Biondini_Chong_Kevrekidis_2025,https://doi.org/10.1111/sapm.70190,PhysRevE.95.062216,CHONG2024103352}, the Toda lattice \cite{BIONDINI2024134315, PhysRevA.24.2595}, and discrete conservation laws \cite{YANG2026103695, Sprenger_2024, CHONG2022133533}. Recently, DSW is numerically shown to exist also in a two-dimensional (in space) model which is the Kadomtsev-Petviashvili (KP) equation \cite{cdvf-xnfw}. Besides the numerical existence, the DSW is also experimentally discovered \cite{PhysRevLett.120.194101,PhysRevE.75.021304,PhysRevE.80.056602}. The theoretical analysis of this particular dispersive wave is performed on the basis of the so-called Whitham modulation theory \cite{whitham2011linear,https://doi.org/10.1111/sapm.12651, Abeya_2023, Ablowitz_2018,PhysRevE.96.032225} which leads to a system of partial differential equations (PDEs) governing the closed dynamics of the slowly varying parameters of the periodic traveling waves of the dispersive models of interest. Interestingly, the Whitham modulation system admits two very useful reductions at both the harmonic and soliton limits \cite{10.1063/1.1947120}. Such reductions encode important information about distinct edge characteristics of the DSW. The rarefaction wave (RW), on the other hand, is a simple wave structure \cite{carretero2024nonlinear, chong2018coherent}. Specifically, it admits numerical emergence also in dispersive media. This particular wave pattern can be described and approximated by the self-similar solutions \cite{EL201611} associated with the dispersionless limit of the dispersive model.

In the present work, we shall focus on a dispersive discrete lattice \cite{CHEN2023133652}, the so-called discrete modified KdV equation, stemming from the continuum modified Korteweg-De Vries (KdV) equation \cite{Chen_2019,Chen_2018,doi:10.1137/21M1465329} which is an integrable system. In particular, we study, both analytically and numerically, the rarefaction and dispersive shock waves of such discrete system by proposing relevant quasi-continuum models, via the analytical methods specified in Ref.~\cite{CHONG2022133533}, to approximate different edge features including the edge speeds and wavenumber of the discrete DSW of the lattice. 

This paper is structured as follows. In Section \ref{sec: Intro and model description}, we give an overview of the target model which shall be the main focus of this work. In particular, we discuss some preliminaries and facts related to such model and simultaneously introduce the definition of the so-called Riemann problem associated with the model. Next, we derive relevant quasi-continuum models in Section \ref{sec: derivation of quasi-continuum models}, which will be applied for approximating the two dispersive wave structures of the target model. In Section \ref{sec: periodic solutions}, we compute and analyze the periodic traveling-wave solutions of each quasi-continuum model derived in Section \ref{sec: derivation of quasi-continuum models}. In Sections \ref{sec: conservation laws} and \ref{sec: Whitham analysis}, we list some necessary conservation laws of each quasi-continuum model, which serve as one particular prerequisite for performing the Whitham analysis. Then, we derive the Whitham modulation equations by the method of averaging the conservation laws. Based on the Whitham modulation equations, we perform two important reductions in Section \ref{sec: Reduc of Whitham systems}, which serve as the derivation of the ``DSW-fitting" method. We then perform the DSW fitting to make analytical predictions on multiple edge characteristics of the DSWs of the quasi-continuum models in Section \ref{sec: DSW fitting}. In Section \ref{sec: Numerical methods}, we discuss the relevant numerical methods utilized in solving each quasi-continuum model and the appropriate way to construct a consistent set of initial conditions for the purpose of numerical comparison that shall be done in the later Section \ref{sec: numerical vali}. In Section \ref{sec: RWs}, we compute the analytical self-similar solutions corresponding to each quasi-continuum model which shall be used to approximate the numerical RW of the discrete modified KdV equation. We then make some necessary numerical comparisons in Section \ref{sec: numerical vali} to examine the performance of all the quasi-continuum models in approximating the different edge features of the discrete DSW of the lattice. Finally, the paper ends in the conclusion section of \ref{sec: conclusions} with some open questions related to some potential directions for future research.

\section{Model description and theoretical setup}\label{sec: Intro and model description}

We consider the integrable discrete modified KdV (dmKdV) equation \cite{CHEN2023133652} in the following normalized form:
\begin{equation}\label{e: dmKdV equation}
    \frac{du_n}{dt} = \left(1+u_n^2\right)\left(u_{n+1} - u_{n-1}\right),
\end{equation}
where $n \in \mathbb{Z}$ denotes the lattice site.

We notice that the discrete lattice in Eq.~\eqref{e: dmKdV equation} is dispersive. To see this, we can compute the linear dispersion relation of Eq.~\eqref{e: dmKdV equation} by looking for a plane-wave solution in the form of an infinitesimal perturbation near the homogeneous state of $\overline{u}$: $u_n(t) = \overline{u} + a\exp\left[i(kn-\omega t)\right]$, where $0<a\ll 1$ is a small parameter. Substitution of this plane-wave ansatz into Eq.~\eqref{e: dmKdV equation} yields the following linear dispersion relation, upon the elimination of $a$,
\begin{equation}\label{e: ldr of dmKdV}
    \omega(\overline{u},k) = -2(1+\overline{u}^2)\sin(k).
\end{equation}
We notice that $\partial_{kk}\omega \neq 0$ for every wavenumber $k$, so this demonstrates that the discrete mKdV equation \eqref{e: dmKdV equation} is dispersive.

In the present work, we study the nonlinear dispersive wave patterns numerically observed from the Riemann problem associated with Eq.~\eqref{e: dmKdV equation}. Namely, we consider the Cauchy problem of Eq.~\eqref{e: dmKdV equation} with the following Riemann initial condition
\begin{equation}\label{Riemann ICs}
    u_n(0) = \begin{cases}
        u_-, \quad n \leq 0, \\
        u_+, \quad n > 0,
    \end{cases}
\end{equation}
where $u_{\pm} \in \mathbb{R}$ refer to the two backgrounds of the Riemann initial data. 

\begin{figure}[b!]
    \centering
    \includegraphics[width=1.05\linewidth]{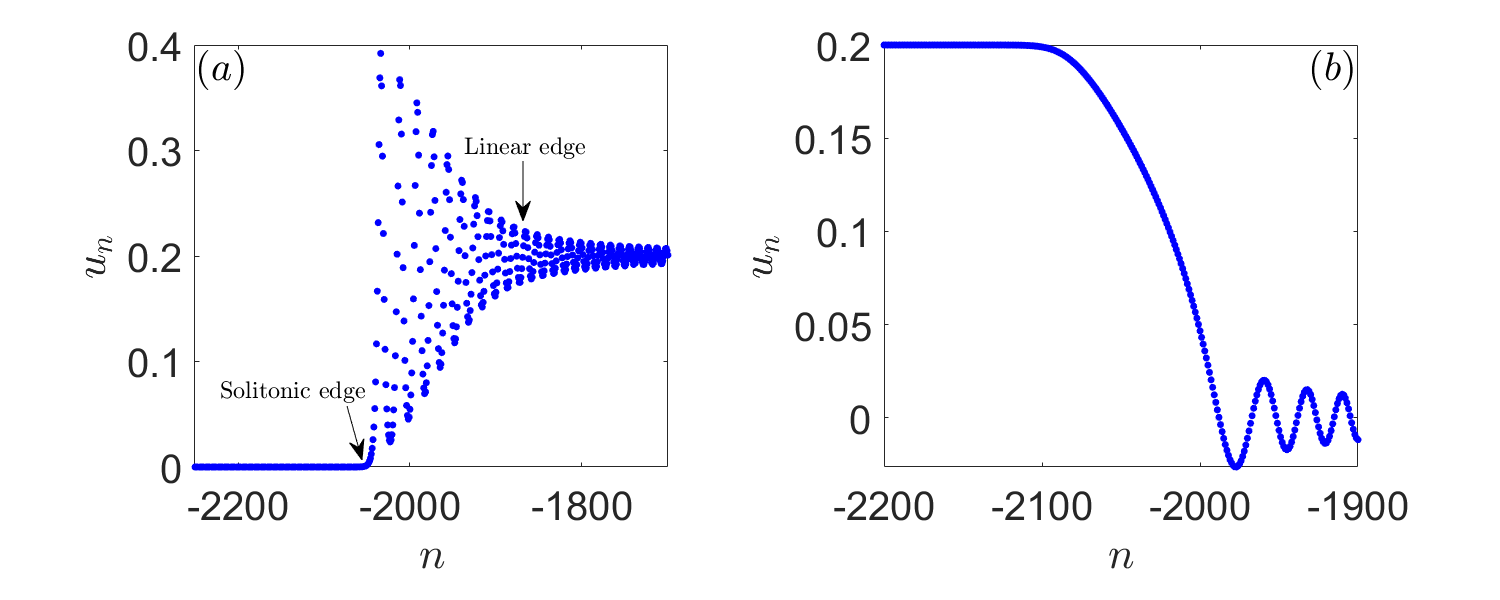}
    \caption{The evolution of the Riemann problem associated with the discrete mKdV equation \eqref{e: dmKdV equation}. The dynamics is shown at $t = 1000$ with the backgrounds of $u_- = 0$ and $u_+ = 0.2$ for the panel $(a)$, while $u_- = 0.2$ and $u_+ = 0$ for $(b)$.} 
    \label{fig:Riemann problem associated with lattice}
\end{figure}

We shall see from a specific numerical simulation that when $u_- < u_+$, we obtain a numerical discrete dispersive shock wave, while a rarefaction wave when $u_- > u_+$. Fig.~\ref{fig:Riemann problem associated with lattice} shows the dynamics of the Riemann problem associated with Eq.~\eqref{e: dmKdV equation}. In particular, note that when $u_- = 0$, $u_+ = 0.2$, the panel $(a)$ shows the spatial profile of a dispersive shock wave at $t = 1000$, where its linear and solitonic edges are labeled. On the other hand, the panel $(b)$ shows that of a numerical rarefaction wave of the lattice \eqref{e: dmKdV equation}. Both waves are propagating to the left. In the later section, we shall make theoretical predictions on these speeds of propagation including the DSW solitonic and linear-edge speeds.

\section{Quasi-continuum approximations}\label{sec: derivation of quasi-continuum models}

To derive some relevant quasi-continuum models for Eq.~\eqref{e: dmKdV equation}, we first consider the following multi-scale change of variables \cite{CHEN2023133652}:
\begin{equation}\label{e: multi-scale ansatz}
   \begin{aligned}
    &u_n(t) = \epsilon u(\xi,\tau);\\
    &\xi = \epsilon(n+2t), \quad \tau = \frac{1}{3}\epsilon^3t,
    \end{aligned}
\end{equation}
where $0<\epsilon\ll 1$ is a formal smallness parameter.

Substituting the ansatz in Eq.~\eqref{e: multi-scale ansatz} into the dmKdV equation \eqref{e: dmKdV equation} yields the following continuum modified KdV (mKdV) equation, by collecting relevant terms up to the order of $\mathcal{O}(\epsilon^4)$:
\begin{equation}\label{e: continuum mKdV}
    u_\tau = 6u^2u_\xi + u_{\xi\xi\xi}.
\end{equation}
Looking for a plane-wave solution in the form of $u(\xi,\tau) = \overline{u} + a\exp\left[i(K\xi - \Omega \tau)\right]$ ($0 < a \ll1$) and substitution into Eq.~\eqref{e: continuum mKdV} yields the following linear dispersion relation by eliminating the small parameter of $a$:
\begin{equation}\label{ldr of continuum mKdV}
    \Omega_{\text{mKdV}}\left(\overline{u},K\right) = -6\overline{u}^2K + K^3.
\end{equation}

Instead of the quasi-continuum mKdV model in Eq.~\eqref{e: continuum mKdV}, we can instead define the slow spatial and temporal variables in the following manner.
\begin{equation}\label{eq: second set of slow variables}
   \begin{aligned}
    &u_n(t) = u(X,T),\\
    &X = \epsilon n, \quad T = \epsilon t.
    \end{aligned}
\end{equation}
Based on the definitions of the slow spatial and temporal variables of $X$ and $T$, this immediately implies the following relations between the wavenumbers and frequencies.
\begin{equation}\label{Relations of wavenumbers and freq}
    K = k/\epsilon, \quad \Omega = \omega / \epsilon.
\end{equation}

Then, substitution of the ansatz in Eq.~\eqref{eq: second set of slow variables} into the discrete mKdV \eqref{e: dmKdV equation} yields, by collecting relevant terms up to the order of $\mathcal{O}(\epsilon^3)$,
\begin{equation}\label{e: second continuum model}
    u_T = \left(1+u^2\right)\left(2u_X + \frac{\epsilon^2}{3}u_{XXX}\right).
\end{equation}
We then look for a plane-wave solution in the form of $u(X,T) = \overline{u} + a\exp\left[i(KX-\Omega T)\right]$ ($0 < a \ll 1$) to arrive at the following linear dispersion relation for Eq.~\eqref{e: second continuum model}, upon the substitution of such ansatz into Eq.~\eqref{e: second continuum model}.
\begin{equation}\label{e: unbounded ldr of second model}
    \Omega_{\text{n}}(\overline{u},K) = -\left(1+\overline{u}^2\right)\left(2K-\frac{\epsilon^2}{3}K^3\right).
\end{equation}
It can be readily seen from Eq.~\eqref{e: unbounded ldr of second model} that the proposed model in Eq.~\eqref{e: second continuum model} does not have a bounded linear dispersion relation. Namely, $\Omega_{\text{n}} \to \pm\infty$ as $K \to \pm \infty$. In order to obtain a quasi-continuum model with bounded dispersion relation, we rewrite Eq.~\eqref{e: second continuum model} as follows.
\begin{equation}\label{e: rewrite second model}
    \arctan(u)_T = 2\left(1+\frac{\epsilon^2}{6}\partial^2_X\right)u_X.
\end{equation}
Then, we invert the operator $1+\frac{\epsilon^2}{6}\partial^2_X$ on the right-hand side of Eq.~\eqref{e: rewrite second model} to obtain, by performing Taylor expansion and collecting terms up to the order of $\mathcal{O}(\epsilon^2)$:
\begin{equation}\label{e: regularized model}
    \arctan(u)_T - \frac{\epsilon^2}{6}\arctan(u)_{XXT} = 2u_X.
\end{equation}
The linear dispersion relation of the ``regularized" model in Eq.~\eqref{e: regularized model} admits the following linear dispersion relation:
\begin{equation}\label{ldr of the regularized model}
    \Omega_{\text{r}}(\overline{u},K) = -\frac{2\left(1+\overline{u}^2\right)K}{1 + \frac{\epsilon^2K^2}{6}},
\end{equation}
which is clearly bounded as $\Omega_{\text{r}} \to 0$ as $K \to \pm\infty$.

\begin{figure}[t!]
    \centering
    \includegraphics[width=0.525\linewidth]{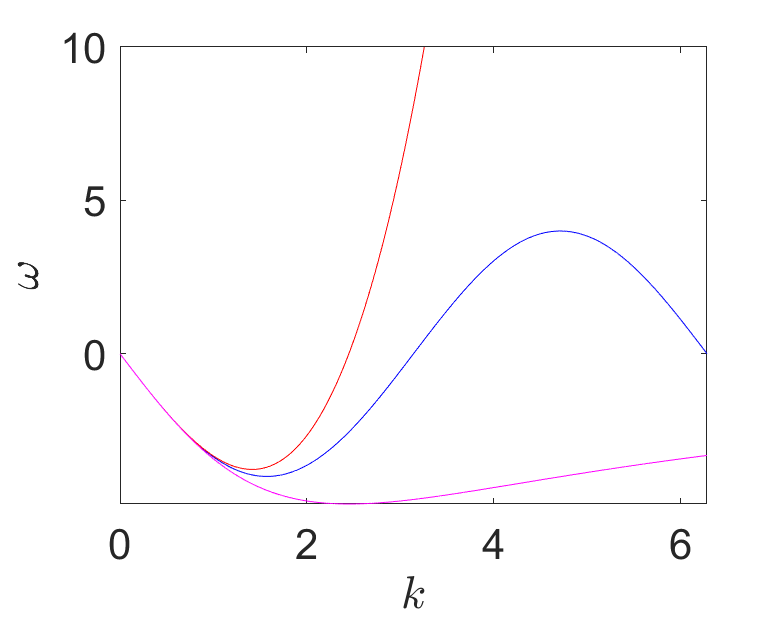}
    \caption{The comparison of the linear dispersion relations of the three models in Eqs.~\eqref{e: dmKdV equation} (blue), \eqref{e: second continuum model} (red) and \eqref{e: regularized model} (magenta).}
    \label{fig:ldr comparison}
\end{figure}

Fig.~\ref{fig:ldr comparison} depicts the comparison of the three linear dispersion relations in Eqs.~\eqref{e: ldr of dmKdV}, \eqref{e: unbounded ldr of second model} and \eqref{ldr of the regularized model}. In particular, we can see clearly that the dispersion relation of the non-regularized model in Eq.~\eqref{e: second continuum model}, represented by the red curve, asymptotes to positive infinity drastically as the wavenumber $k$ increases, while that of the regularized model \eqref{e: regularized model}, described via the magenta curve, remains bounded and approaches to zero as $k\to\infty$.

We shall apply the three quasi-continuum models in Eqs.~\eqref{e: continuum mKdV}, \eqref{e: second continuum model} and \eqref{e: regularized model} to make approximations on the nonlinear dispersive waves numerically observed in the discrete mKdV equation \eqref{e: dmKdV equation}. Finally, we end this section with a brief discussion on one potential limitation of these quasi-continuum models (e.g. \eqref{e: continuum mKdV}). In particular, we expect these quasi-continuum models to provide reasonable approximations on the discrete dispersive wave structures of Eq.~\eqref{e: dmKdV equation} provided that their scales, denoted as $\Gamma$, are considerably larger than the spacing of the lattice, which is proportional to the formal smallness parameter of $\epsilon$ \cite{PEYRARD198488}. Namely, only when $\Gamma \gg \epsilon$, it can then be expected that the three quasi-continuum models in Eqs.~\eqref{e: continuum mKdV}, \eqref{eq: second set of slow variables} and \eqref{e: regularized model} can yield reasonable approximated dispersive wave patterns for these of the lattice \eqref{e: dmKdV equation}. As a result, we shall not utilize them for any early-time dynamics comparison with Eq.~\eqref{e: dmKdV equation}.

\section{Periodic traveling-wave solutions}\label{sec: periodic solutions}

In this section, we seek the periodic traveling-wave solutions for the discrete lattice \eqref{e: dmKdV equation} and the three quasi-continuum models in Eqs.~\eqref{e: continuum mKdV}, \eqref{e: second continuum model} and \eqref{e: regularized model}. These periodic waves play an essential role and serve as one of the necessary prerequisites for the Whitham modulation theory which we shall discuss in Section \ref{sec: Whitham analysis}.

Firstly, for the discrete lattice in Eq.~\eqref{e: dmKdV equation}, we review its periodic traveling-wave solution described in \cite{CHEN2023133652}. Specifically, the discrete mKdV equation \eqref{e: dmKdV equation} admits two family of periodic traveling waves. On the one hand, it has the following so-called ``dnoidal" periodic wave solution expressed in terms of the Jacobi elliptic functions:
\begin{equation}\label{dnoidal waves}
   \begin{aligned}
    &u_n(t) = \frac{\text{sn}\left(\alpha; m\right)}{\text{cn}\left(\alpha;m\right)}\text{dn}\left(\alpha n + ct; m\right),\\
    &c = \frac{2\text{sn}\left(\alpha;m\right)}{\text{cn}\left(\alpha;m\right)}, 
    \end{aligned}
\end{equation}
where $m \in (0,1)$ denotes the elliptic modulus.

On the other hand, the lattice \eqref{e: dmKdV equation} also possesses the following ``cnoidal" periodic wave solution:
\begin{equation}\label{cnoidal waves}
   \begin{aligned}
   &u_n(t) = m\frac{\text{sn}\left(\alpha;m\right)}{\text{dn}\left(\alpha;m\right)}\text{cn}\left(\alpha n+ ct; m\right),\\
    &c = \frac{2\text{sn}\left(\alpha;m\right)}{\text{dn}\left(\alpha;m\right)}.
    \end{aligned}
\end{equation}
At the soliton limit where $m \to 1$, it has been shown that the two periodic waves in Eqs.~\eqref{dnoidal waves} and \eqref{cnoidal waves} shall asymptote to the same solitary-wave solutions \cite{CHEN2023133652}:
\begin{equation}\label{solitary-wave soln for lattice}
    u_n(t) = \sinh(\alpha)\sech(\alpha n+ ct), \quad c = 2\sinh(\alpha).
\end{equation}

Next, for the continuum mKdV equation \eqref{e: continuum mKdV}, we look for the periodic solutions in the form of 
\begin{equation}\label{e: TW ansatz for continuum mKdV}
    u(\xi,\tau) = u(z), \quad z = \alpha \xi + c\tau,
\end{equation}
where $\alpha > 0 $ is a constant and $c \in \mathbb{R}$ refers to the speed of propagation of the waves.

Substitution of the ansatz \eqref{e: TW ansatz for continuum mKdV} into Eq.~\eqref{e: continuum mKdV} yields the following co-traveling frame ordinary differential equation (ODE), upon integration with respect to $z$ twice.
\begin{equation}\label{e: co-traveling ode of mKdV}
    \alpha^3(u_z)^2 = cu^2 - \alpha u^4 - 2Au - B,
\end{equation}
where $A,B$ are two constants of integration.

In particular, based on the co-traveling frame ODE in Eq.~\eqref{e: co-traveling ode of mKdV}, we can compute the solitary-wave solution by assuming that $u(z)\to0$ as $z \to \pm \infty$. This implies that $A = B = 0$, and then we can perform direct integration of Eq.~\eqref{e: co-traveling ode of mKdV} by the method of quadrature to arrive at
\begin{equation}\label{solitary-wave solution to mKdV}
    u(\xi,\tau) = \sqrt{\frac{c}{\alpha}}\sech\left(\frac{\sqrt{c}}{\alpha^{3/2}}\left(\alpha\xi+c\tau-\xi_0\right)\right),
\end{equation}
where $\xi_0$ denotes an arbitrary phase parameter.

Hence, the soliton amplitude-speed relation is described as
\begin{equation}\label{soliton amplitude-speed relation}
    a = \sqrt{\frac{c}{\alpha}},
\end{equation}
where $a$ is the amplitude of the soliton in Eq.~\eqref{solitary-wave solution to mKdV}.

Moreover, the periodic solutions associated with the model Eq.~\eqref{e: continuum mKdV} can be similarly obtained by integrating Eq.~\eqref{e: co-traveling ode of mKdV}, however, without the assumption that $A = B = 0$, and the solutions can be expressed, in principle, in terms of Jacobi elliptic functions \cite{doi:10.1137/15M1015650}. We shall omit the derivation of the periodic solutions here for brevity since they are out of the considerations of this present work.

Next, for the other two quasi-continuum models in Eqs.~\eqref{e: second continuum model} and \eqref{e: regularized model}, we assume the following traveling-wave ansatz,
\begin{equation}\label{TW ansatz for other two models}
    u(X,T) = u(\tilde{z}), \quad \tilde{z} = \alpha X + cT.
\end{equation}
Substitution of the ansatz \eqref{TW ansatz for other two models} into the two models in Eqs.~\eqref{e: second continuum model} and \eqref{e: regularized model} yields.
\begin{equation}\label{e: co-traveling ODEs for other two models}
    \begin{aligned}
        &\frac{\epsilon^2\alpha^3}{3}(u_{\tilde{z}})^2 = 2c\left(u\arctan(u) - \frac{1}{2}\log\left(1+u^2\right)\right) \\
        &- 2\alpha u^2 - 2Au - B,\\
        &\frac{\epsilon^2\alpha^2c}{6}(\tilde{u}_{\tilde{z}})^2 = c\tilde{u}^2 - 4\alpha\log\left|\sec(\tilde{u})\right| - 2A\tilde{u} - B,
    \end{aligned}
\end{equation}
respectively, where $A,B$ are two constants of integration, and $\tilde{u} \equiv \arctan(u)$.

To the best of our knowledge, both the solitary-wave and periodic traveling-wave solutions do not admit analytical expressions for both models in Eqs.~\eqref{e: second continuum model} and \eqref{e: regularized model}. Nevertheless, we can still show the existence of such solutions by performing a relevant phase-plane analysis via visualizing the potential curves associated with the two co-traveling frame ODEs in Eq.~\eqref{e: co-traveling ODEs for other two models}. We shall draw the potential curves of the two co-traveling frame ODEs in Eq.~\eqref{e: co-traveling ODEs for other two models} in the following forms
\begin{equation}\label{potential curves}
   \begin{aligned}
    &P_{\text{n}}(u) = \frac{3}{\epsilon^2\alpha^3}\bigg[-2c\left(u\arctan(u)-\frac{1}{2}\log(1+u^2)\right) \\
    &+ 2\alpha u^2 + 2Au + B\bigg],\\
    &P_{\text{r}}(\tilde{u}) = \frac{6}{\epsilon^2\alpha^2c}\left(-c\tilde{u}^2+4\alpha\log\left|\sec(\tilde{u})\right| + 2A\tilde{u} + B\right).
    \end{aligned}
\end{equation}

\begin{figure}[t!]
    \centering
    \includegraphics[width=1.05\linewidth]{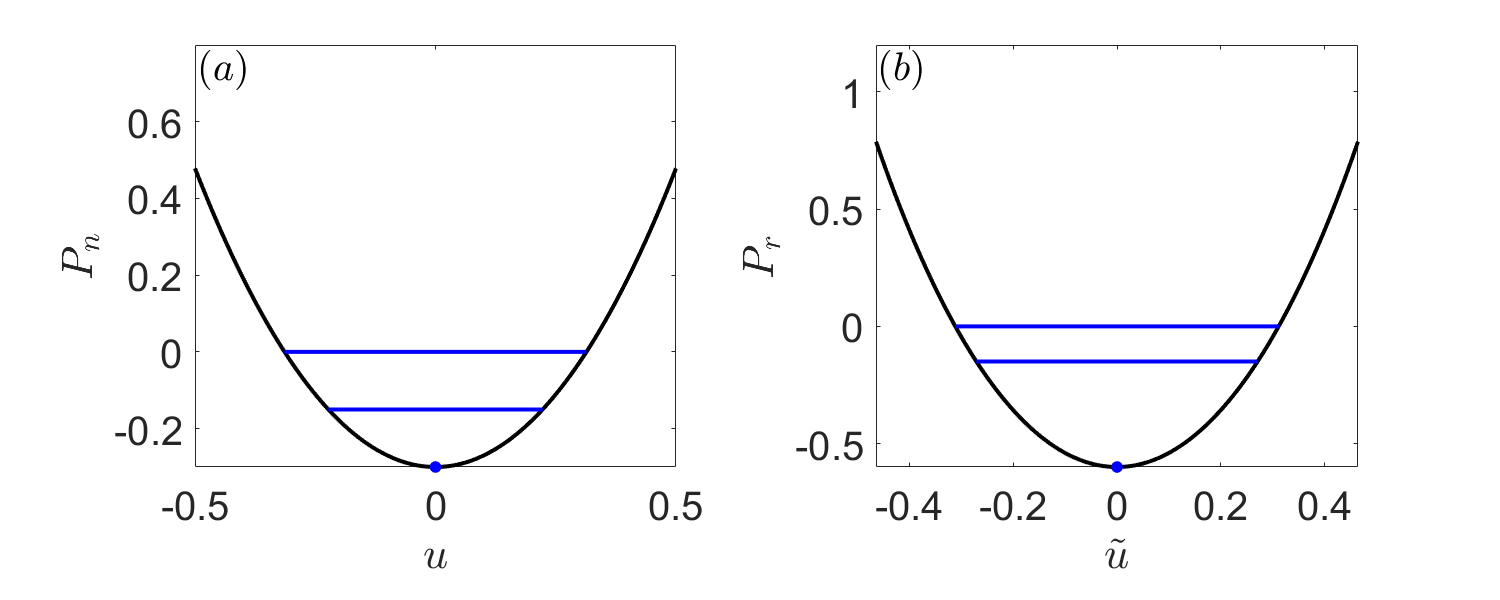}
    \caption{The potential curves in Eq.~\ref{potential curves}. The panels $(a)$ and $(b)$ depict the potential curve of $P_{\text{n}}(u)$ and $P_{\text{r}}(\tilde{u})$, respectively. }
    \label{fig:Potential curves}
\end{figure}

Fig.~\ref{fig:Potential curves} display the potential curves in Eq.~\eqref{potential curves}. We notice that the two blue horizontal lines in each panel demonstrate the associated periodic orbits of the co-traveling frame ODEs which are the periodic waves corresponding to the original models in Eqs.~\eqref{e: second continuum model} and \eqref{e: regularized model}.

\section{Conservation laws}\label{sec: conservation laws}

In this section, we list some relevant conservation laws associated with each quasi-continuum model. These conservation laws shall be an indispensable prerequisite for performing Whitham analysis. Specifically, to obtain useful insights on the slowly varying dynamics of the parameters of the periodic traveling waves, we need to derive the so-called Whitham modulation system which is described by a system of partial differential equations, called the Whitham modulation equations, governing a closed system of dynamics of all the relevant parameters. The modulation equations can be derived by various approaches \cite{EL201611}. However, we shall mainly rely on the method of ``averaging the conservation laws", and hence the necessity of knowing relevant conservation laws of the three quasi-continuum models in Eqs.~\eqref{e: continuum mKdV}, \eqref{e: second continuum model} and \eqref{e: regularized model}. In addition, we notice that since the co-traveling frame ODEs for all the quasi-continuum models have an order of three, this suggests that the associated periodic traveling waves are three-parameter family of solutions. In order to construct a closed slowly varying dynamics of these three parameters, we need only two conservation laws which shall yield two modulation equations upon performing averaging operation. We shall introduce the remaining modulation equation in Section \ref{sec: Whitham analysis}, which is the so-called ``conservation of waves".

Firstly, the model \eqref{e: continuum mKdV} possesses the following two conservation laws:
\begin{equation}\label{e: conservation laws of 1st model}
   \begin{aligned}
    &u_\tau = 2(u^3)_\xi + u_{\xi\xi\xi},\\
    &\frac{1}{2}(u^2)_\tau = \frac{3}{2}(u^4)_\xi + (uu_{\xi\xi})_\xi - \frac{1}{2}\left[\left(u_\xi\right)^2\right]_\xi,
    \end{aligned}
\end{equation}
where the first and second equation in \eqref{e: conservation laws of 1st model} refer to the conservation of mass and linear momentum, respectively.

The other two quasi-continuum models in Eqs.~\eqref{e: second continuum model} and \eqref{e: regularized model} admit the following two conservation laws.
For Eq.~\eqref{e: second continuum model},
\begin{equation}\label{e: conservation laws of 2nd model}
    \begin{aligned}
        &\arctan(u)_T = 2u_X + \frac{\epsilon^2}{3}u_{XXX},\\
        &\frac{1}{2}\log\left(1+u^2\right)_T = (u^2)_X + \frac{\epsilon^2}{3}\left(uu_{XX}\right)_X - \frac{\epsilon^2}{6}\left[\left(u_X\right)^2\right]_X.
    \end{aligned}
\end{equation}
Then, for Eq.~\eqref{e: regularized model},
\begin{equation}\label{e: conservation laws of 3rd model}
    \begin{aligned}
    &\left(\tilde{u} - \frac{\epsilon^2}{6}\tilde{u}_{XX}\right)_T - 2\tan\left(\tilde{u}\right)_X = 0,\\
    &\frac{1}{2}(\tilde{u}^2)_T + \frac{\epsilon^2}{12}\left[(\tilde{u}_X)^2\right]_T \\
    &- \frac{\epsilon^2}{6}\left(\tilde{u}\tilde{u}_{XT}\right)_X - 2\left(\tilde{u}\tan(\tilde{u}) + \log\left|\cos(\tilde{u})\right|\right)_X = 0,
    \end{aligned}
\end{equation}
where we recall that $\tilde{u} = \arctan(u)$.

Analogously, for the conservation laws in Eqs.~\eqref{e: conservation laws of 2nd model} and \eqref{e: conservation laws of 3rd model}, they represent the mass and momentum conservation, respectively.

\section{Whitham modulation equations}\label{sec: Whitham analysis}

In this section, we derive the Whitham modulation equations which describe the closed slowly varying dynamics of all the relevant parameters of the periodic traveling waves revisited in Section \ref{sec: periodic solutions}. 

Firstly, regarding the modulation system associated with the modified KdV equation \eqref{e: continuum mKdV}, it is well documented in, for example, \cite{doi:10.1137/15M1015650}, so we shall not re-derive similar results in the present work for brevity. We now attempt to compute the Whitham modulation equations for the two quasi-continuum models in Eqs.~\eqref{e: second continuum model} and \eqref{e: regularized model} by applying the method of averaging the conservation laws which are specified in Section \ref{sec: conservation laws}. We introduce first some necessary preliminaries for applying such method. In particular, we first define the following fast phase $\theta$:
\begin{equation}\label{fast-phase theta}
    \theta = \frac{KX - \Omega T}{\epsilon},
\end{equation}
and then we look for periodic traveling waves associated with the models of Eqs.~\eqref{e: second continuum model} and \eqref{e: regularized model} in the following multiple-scale form.
\begin{equation}\label{multi-scale expansions of periodic solutions}
    u(X,T) = \phi\left(\theta;X,T\right) + \epsilon u_1\left(\theta;X,T\right) + \mathcal{O}\left(\epsilon^2\right),
\end{equation}
where functions of $\varphi\left(\theta;X,T\right)$ and $u_1\left(\theta;X,T\right)$ refer to the periodic traveling-wave solutions of the models with fixed period of $2\pi$. Notice that the period of $2\pi$ is a convention, and it can be arbitrary. Nevertheless, it is important to note that the periodic traveling waves ought to have a fixed period as this is one particular prerequisite \cite{EL201611} of performing Whitham averaging for obtaining the modulation equations. In addition, we note that the exact analytical expression of $\varphi(\theta;X,T)$ is not known for either model of \eqref{e: second continuum model} or \eqref{e: regularized model}. However, we have proven their existences via the relevant phase-plane analysis in Section \ref{sec: periodic solutions}, and we make the necessary assumptions that their periods can still be fixed through the process of reparametrization \cite{Yang_Biondini_Chong_Kevrekidis_2025}. Next, the fast phase of $\theta$ is further defined according to 
\begin{equation}
    \theta_X = K/\epsilon, \quad \theta_T = -\Omega/\epsilon,
\end{equation}
where both the wavenumber $K$ and frequency $\Omega$ are functions of the two slow variables of $X$ and $T$. Then, the compatibility condition of $\theta_{XT} = \theta_{TX}$ leads to the following equation.
\begin{equation}
    K_T + \Omega_X = 0,
\end{equation}
which is the so-called ``conservation of waves", and is the first modulation equation of the full closed modulation system.

We now define the following averaging operator: For a given function $F$,
\begin{equation}
    \overline{F(\phi)} = \frac{1}{2\pi}\int_0^{2\pi}F(\phi(\theta))d\theta.
\end{equation}
Then, applying such averaging operation to the conservation laws specified in Section \ref{sec: conservation laws} and collecting relevant terms up to the order of $\mathcal{O}(\epsilon)$ yields the closed system of modulation equations.

On the one hand, the Whitham modulation equations for Eq.~\eqref{e: second continuum model} are:
\begin{equation}\label{e: modulation equations for 2nd model}
   \begin{aligned}
    &K_T + \Omega_X = 0,\\
    &\overline{\arctan(\phi)}_T - 2\overline{\phi}_X = 0,\\
    &\frac{1}{2}\overline{\log\left(1+\phi^2\right)}_T - \left(\overline{\phi^2}+\frac{K^2}{2}\overline{\left(\phi_\theta\right)^2}\right)_X = 0.
    \end{aligned}
\end{equation}
On the other hand, the modulation system associated with the model \eqref{e: regularized model} reads
\begin{equation}\label{eq: modulation equations for 3rd model}
   \begin{aligned}
    &K_T + \Omega_X = 0,\\
    &\overline{\tilde{\phi}}_T - 2\overline{\tan\left(\tilde{\phi}\right)}_X = 0,\\
    &\frac{1}{2}\overline{\tilde{\phi}^2}_T + \frac{1}{12}\left(K^2\overline{\left(\tilde{\phi}_\theta\right)^2}\right)_T \\
    &-\frac{1}{6}\left(K\Omega\overline{\left(\tilde{\phi}_\theta\right)^2}\right)_X - 2\left(\overline{\tilde{\phi}\tan(\tilde{\phi}) + \log\left|\cos(\tilde{\phi})\right|}\right)_X = 0,
    \end{aligned}
\end{equation}
where $\tilde{\phi} \equiv \tan(\phi)$.

\section{Reduction of the Whitham system}\label{sec: Reduc of Whitham systems}

The Whitham modulation equations presented in the form of Eqs.~\eqref{e: modulation equations for 2nd model} and \eqref{eq: modulation equations for 3rd model} do not generally have practical application. Nonetheless, it can be very useful to perform a suitable set of reduction of the two Whitham systems at both the harmonic and solitonic limits, since these limit modulation systems shall encode important insights on distinct edge features of the DSWs. 

We demonstrate the precise derivation of such reduction only for the Whitham system in Eq.~\eqref{e: modulation equations for 2nd model} since it is completely analogous for that of Eq.~\eqref{eq: modulation equations for 3rd model}. Firstly, at the harmonic edge, it is expected that the wave amplitude, denoted as $a$, to asymptote to zero (i.e. $a \to 0$) and hence the waves admits small amplitudes. This further implies the following fact \cite{10.1063/1.1947120}: For a given function $F$,
\begin{equation}
    \overline{F(\phi)} \to F(\overline{\phi}), \quad \overline{\left(\phi_\theta\right)^2} \to 0.
\end{equation}
Therefore, the modulation system \eqref{e: modulation equations for 2nd model} degenerates into the following two equations.
\begin{equation}\label{e: degenerated Whitham system}
   \begin{aligned}
    &K_T + \Omega_{\text{n}X} = 0,\\
    &\overline{\phi}_T - 2\left(1+\overline{\phi}^2\right)\overline{\phi}_X = 0,
    \end{aligned}
\end{equation}
where $\Omega_{\text{n}}$ refers to the linear dispersion relation in Eq.~\eqref{e: unbounded ldr of second model}.

Then, we can cast the system Eq.~\eqref{e: degenerated Whitham system} into the following characteristic form, upon the utilization of the chain rules:
\begin{equation}\label{e: char form}
    \begin{bmatrix}
        K \\
        \overline{\phi}
    \end{bmatrix}_T + \begin{bmatrix}
        \partial_K\Omega_{\text{n}} & \partial_{\overline{\phi}}\Omega_{\text{n}} \\
        0 & -2\left(1+\overline{\phi}^2\right)
    \end{bmatrix}
    \begin{bmatrix}
        K \\
        \overline{\phi}
    \end{bmatrix}_X = 0.
\end{equation}
Then, we notice that the Jacobian matrix in Eq.~\eqref{e: char form} admits the following left eigenpare of $(\lambda_1,v_1)$:
\begin{equation}\label{left-eigenpare}
    \lambda_1 = \partial_K\Omega_{\text{n}},\quad v_1 = 
    \begin{bmatrix}
    2(1+\overline{\phi}^2)+\partial_K\Omega_{\text{n}} &\partial_{\overline{\phi}}\Omega_{\text{n}}
    \end{bmatrix}.
\end{equation}
We multiply the characteristic equation \eqref{e: char form} with the left eigenvector of $v_1$ in Eq.~\eqref{left-eigenpare} to obtain that, by chain rules
\begin{equation}\label{linear-edge simple-wave ODE}
    \frac{dK}{d\overline{\phi}} = -\frac{\partial_{\overline{\phi}}\Omega_{\text{n}}}{2\left(1+\overline{\phi}^2\right)+\partial_{K}\Omega_{\text{n}}}.
\end{equation}
The equation \eqref{linear-edge simple-wave ODE} is the so-called ``simple-wave" ordinary differential equation (ODE) at the harmonic limit of the modulation system. In particular, this ODE discloses the important relation between the wavenumber $K$ and the associated averaged background of $\overline{\phi}$. We shall see its application in Section \ref{sec: DSW fitting} as a theoretical approach to make predictions on the different edge features of the DSWs of the models of interested.

A similar simple-wave ODE can be derived also at the solitonic limit where the wavenumber $K \to 0$. In this case, it is more convenient to define the following ``conjugate dispersion relation" \cite{10.1063/1.1947120}, denoted as $\Omega^s_{\text{n}}$:
\begin{equation}
    \Omega^s_{\text{n}}\left(\overline{u},\widetilde{K}\right) = -i\Omega_{\text{n}}\left(\overline{u},i\widetilde{K}\right),
\end{equation}
where $\widetilde{K}$ denotes the ``conjugate wavenumber". 

After analogous manipulation for the reduction of the modulation system \eqref{e: modulation equations for 2nd model} at the solitonic limit $K \to 0$, we arrive at the following system
\begin{equation}\label{e: simple-wave ode at solitonic limit}
    \frac{d\widetilde{K}}{d\overline{\phi}} = -\frac{\partial_{\overline{\phi}}\Omega^s_{\text{n}}}{2\left(1+\overline{\phi}^2\right) + \partial_{\widetilde{K}}\Omega^s_{\text{n}}}. 
\end{equation}
Moreover, we shall impose some appropriate initial conditions to Eqs.~\eqref{linear-edge simple-wave ODE} and \eqref{e: simple-wave ode at solitonic limit}. On the one hand, at the linear (harmonic) limit where $a \to 0$, it is expected that the wavenumber of $K$ at the solitonic edge of the DSW to be exactly zero so that $K(u_-) = 0$. We need to then integrate this initial-value problem (IVP) from $\overline{\phi} = u_-$ to $\overline{\phi} = u_+$ in order to obtain the wavenumber at the linear edge, denoted as $K_+$, $K_+ = K(u_+)$. On the other hand, regarding the simple-wave ODE at the solitonic limit, we impose the initial condition by the fact that the conjugate wavenumber at the linear edge is zero. Namely, $\widetilde{K}(u_+) = 0$. We then are able to integrate the IVP in Eq.~\eqref{e: simple-wave ode at solitonic limit} up to $\overline{\phi} = u_-$ to compute the conjugate wavenumber at the solitonic edge, denoted as $\widetilde{K}_-$: $\widetilde{K}_- = \widetilde{K}(u_-)$.

\section{DSW fitting}\label{sec: DSW fitting}

We now are ready to solve the simple-wave IVPs derived from Section \ref{sec: Reduc of Whitham systems} to gain important theoretical insights on distinct edge characteristics of the DSWs. We notice that this approach is called the ``DSW-fitting" method, and we can furthermore generalize the method of DSW-fitting as follows. It involves the following two simple-wave IVPs:
\begin{equation}\label{general simple-wave odes}
    \begin{aligned}
        &\frac{dk}{d\overline{u}} = \frac{\partial_{\overline{u}}\omega_0}{V(\overline{u}) - \partial_k\omega_0}, \quad k(u_s) = 0,\\
        &\frac{d\widetilde{k}}{d\overline{u}} = \frac{\partial_{\overline{u}}\omega_s}{V(\overline{u}) - \partial_{\widetilde{k}}\widetilde{\omega}_s}, \quad \widetilde{k}(u_l) = 0,
    \end{aligned}
\end{equation}
where $u_{l,s}$ refer to the values of the backgrounds at the linear and solitonic edge of the DSW, respectively. Moreover, $V(\overline{u})$ represents the dispersionless speed associated with the models of interest. For example, for the continuum mKdV equation \eqref{e: continuum mKdV}, $V(\overline{u}) = -6\overline{u}^2$ as its dispersionless system is given as
\begin{equation}\label{dispersionless form of mKdV equation}
    u_\tau - 6u^2u_\xi = 0.
\end{equation}

Upon obtaining the linear-edge wavenumber $k_l = k(u_l)$ and solitonic-edge conjugate wavenumber $\tilde{k}_s = \tilde{k}(u_s)$, we can further compute the linear-edge and solitonic-edge speeds of the DSW via the group and phase velocities, respectively
\begin{equation}
    \begin{aligned}
        &s_l = \partial_k\omega_0\left(u_l, k_l\right),\\
        &s_s = \frac{\omega_s}{\tilde{k}}\left(u_s,\tilde{k}_s\right).
    \end{aligned}
\end{equation}

We shall perform the DSW fitting to not only the three quasi-continuum models in Eqs.~\eqref{e: continuum mKdV}, \eqref{e: second continuum model} and \eqref{e: regularized model}, but also to the discrete mKdV equation \eqref{e: dmKdV equation}. Firstly, for the dmKdV equation \eqref{e: dmKdV equation} with the linear dispersion relation \eqref{e: ldr of dmKdV}. Solving the IVPs in Eq.~\eqref{general simple-wave odes} yield
\begin{equation}\label{dsw-fitting on lattice}
   \begin{aligned}
    &k_+ = 2\arccos\left(\sqrt{\frac{1+u_-^2}{1+u_+^2}}\right), \quad s_+ = \partial_k\omega_0(u_+,k_+).\\
    &\tilde{k}_- = 2\text{arcsech}\left(\sqrt{\frac{1+u_-^2}{1+u_+^2}}\right), \quad s_- = \frac{\omega_s}{\tilde{k}}\left(u_-,\tilde{k}_-\right),
    \end{aligned}
\end{equation}
where $\omega_0$ is the linear dispersion relation in Eq.~\eqref{e: ldr of dmKdV}. In addition, based on the solitary-wave solution specified in Eq.~\eqref{solitary-wave soln for lattice}, we notice that the soliton amplitude-speed relation is given as
\begin{equation}\label{soliton amp-speed rela 1}
    \left|s_-\right| = \frac{2a_-}{\text{arcsinh}(a_-)},
\end{equation}
where $a_-$ denotes the amplitude of the DSW. So, we are also able to make predictions on the DSW amplitude by solving the nonlinear equation in Eq.~\eqref{soliton amp-speed rela 1} for $a_-$.

Next, for the second model in Eq.~\eqref{e: second continuum model}, solving the simple-wave IVPs yield
\begin{equation}
   \begin{aligned}
    &K_+^{\text{n}} = \frac{\sqrt{6}}{\epsilon}\sqrt{1 - \left(\frac{1+u_-^2}{1+u_+^2}\right)^{2/3}}, \quad S_+^{\text{n}} = \partial_K\Omega_{\text{n}}\left(u_+, K_+^{\text{n}}\right),\\
    &\widetilde{K}_-^{\text{n}} = \frac{\sqrt{6}}{\epsilon}\sqrt{\left(\frac{1+u_+^2}{1+u_-^2}\right)^{2/3} - 1}, \quad S_-^{\text{n}} = \frac{\Omega^s_{\text{n}}}{\widetilde{K}}\left(u_-,\widetilde{K}_-^{\text{n}}\right),
    \end{aligned}
\end{equation}
where $\Omega_{\text{n}}$ refers to the linear dispersion relation \eqref{e: unbounded ldr of second model} of the model \eqref{e: second continuum model}. 

Thirdly, regarding the model in Eq.~\eqref{e: regularized model}, we obtain the following relation between $K$ and $\overline{u}$, by solving the simple-wave IVPs
\begin{equation}\label{transcendal equations}
    \begin{aligned}
       &\frac{\epsilon^2K^2}{12} + \log\left(\frac{\left(6+\epsilon^2K^2\right)\left(1+u_-^2\right)}{6\left(1+\overline{u}^2\right)}\right) = 0,\\
       &-\frac{\epsilon^2\widetilde{K}^2}{12}+\log\left|\frac{\left(-6+\epsilon^2\widetilde{K}^2\right)\left(1+u_+^2\right)}{6\left(1+\overline{u}^2\right)}\right| = 0.
    \end{aligned}
\end{equation}
One shall solve the transcendental equations in Eq.~\eqref{transcendal equations} to obtain the numerical solutions to the linear-edge wavenumber of $K^r_+ = K(u_+)$ and solitonic-edge conjugate wavenumber of $\widetilde{K}^r_- = \widetilde{K}(u_-)$. Moreover, the linear and solitonic-edge speeds of the DSW are given as
\begin{equation}
    S^{\text{r}}_+ = \partial_K\Omega_{\text{r}}\left(u_+, K_+^{\text{r}}\right), \quad S_-^{\text{r}} = \frac{\Omega_{\text{r}}^s}{\widetilde{K}}\left(u_-, \widetilde{K}^{\text{r}}_-\right),
\end{equation}
where $\Omega_{\text{r}}$ is the linear dispersion relation \eqref{ldr of the regularized model} of the regularized model in Eq.~\eqref{e: regularized model}.

Lastly, for the continuum mKdV equation \eqref{e: continuum mKdV}, solving the two IVPs in Eq.~\eqref{general simple-wave odes} yields
\begin{equation}
   \begin{aligned}
    &\widetilde{K}_-^{\text{mKdV}} = K_+^{\text{mKdV}} = 2\sqrt{u_+^2 - u_-^2}, \\
    &S_{+}^{\text{mKdV}} = \partial_K\Omega_{\text{mKdV}}\left(u_+,K_+^{\text{mKdV}}\right), \\
    &S_-^{\text{mKdV}} = \frac{\Omega^s_{\text{mKdV}}}{\widetilde{K}}\left(u_-,\widetilde{K}_-^{\text{mKdV}}\right).
    \end{aligned}
\end{equation}
Also, according to the soliton amplitude-speed relation \eqref{soliton amplitude-speed relation}, the DSW-fitting theoretical prediction on the DSW amplitude, denoted as $a_-$, is as follows.
\begin{equation}\label{DSW-fitting prediction on DSW amplitude of continuum mKdV}
    a_- = \epsilon\sqrt{\left|S_-^{\text{mKdV}}\right|}.
\end{equation}

However, we note that the DSW-fitting theoretical predictions of the continuum mKdV equation \eqref{e: continuum mKdV} are consonant with the $(\xi,\tau)$ coordinates specified in Eq.~\eqref{e: multi-scale ansatz}. Since, as a priori, we shall compare these analytical predictions in the coordinates of $(n,t)$, we need to transform them according to
\begin{equation}\label{Transform back to (n,t)}
    \frac{dn}{dt} = \frac{\epsilon^2}{3}\frac{d\xi}{d\tau} - 2.
\end{equation}

As a priori, we shall compare these DSW-fitting theoretical predictions on the edge speeds of the DSWs with their numerical counterparts in Section \ref{sec: numerical vali}.

\section{Numerical methods and initial conditions}\label{sec: Numerical methods}

In this section, we discuss, in detail, the numerical methods utilized for solving all the relevant models considered in this work. Such discussion is both necessary and important as all of our numerical results discussed in the later sections shall be based on these applied numerical schemes. 

\begin{figure}[t!]
    \centering
    \includegraphics[width=0.55\linewidth]{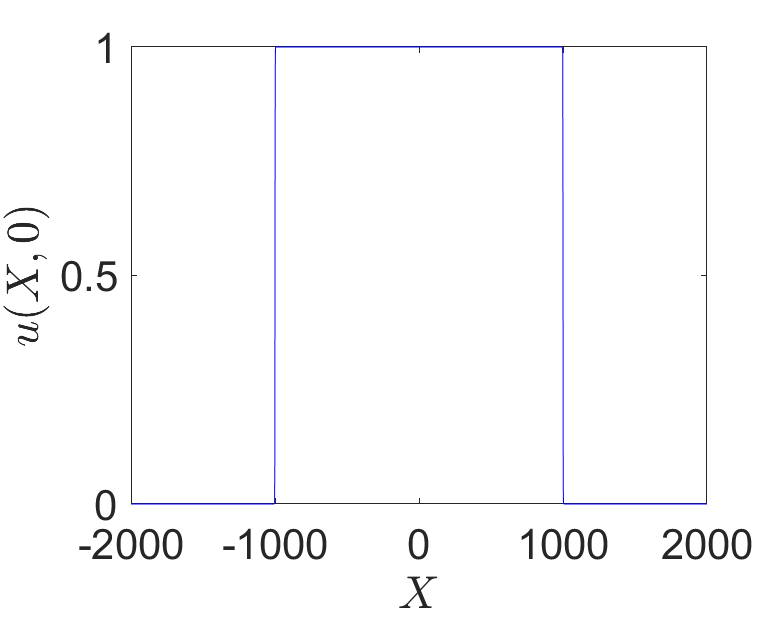}
    \caption{The ``box-type" initial condition with $u_l = 0$, $u_p = 1$ and $\delta = 1$. The two initial jumps are located at $x_l = -1000$ and $x_r = 1000$, respectively.}
    \label{fig:box-type IC}
\end{figure}

Firstly, for the discrete mKdV equation \eqref{e: dmKdV equation}, we apply the standard RK4 time integration scheme to solve the system of difference-difference equations \eqref{e: dmKdV equation}. Then, for the non-regularized and regularized PDEs in Eqs.~\eqref{e: second continuum model} and \eqref{e: regularized model}, we utilize the RK4 time-stepping scheme with the pseudo-spectral method for spatial discritization to numerically solve these two models. Finally, we apply the ETDRK4 time integration scheme \cite{doi:10.1137/S1064827502410633} together with spectral method for spatial discritization to solve the continuum modified KdV equation \eqref{e: continuum mKdV}. We notice that since we utilize the spectral method for spatial discretization, then this shall implicitly assume the periodic boundary conditions. However, it is important to note that the standard Riemann initial condition in Eq.~\eqref{Riemann ICs} clearly does not satisfy the periodic boundary condition. To resolve this issue, we propose the following ``box-type" initial condition for the two continuum models in Eqs.~\eqref{e: second continuum model} and \eqref{e: regularized model}:
\begin{equation}\label{e: IC for non-regularized and regularized models}
   \begin{aligned}
    &u_0(X) = u_l - \frac{1}{2}\left(u_l - u_p\right) \times \\
    &\left[\tanh\left(\delta(X-x_l)\right) - \tanh\left(\delta(X-x_r)\right)\right],
    \end{aligned}
\end{equation}
where $u_0(X) \equiv u(X,0)$, $u_{l,p}$ are two constants determining the upper and lower background values, $\delta > 0$ is the parameter controlling the smoothness of the ``jump", where a larger $\delta$ always suggests a shaper jump/transition of the Riemann initial data. Finally, the values of $x_{l,r}$ represent the locations of the left and right initial jump, respectively. Fig.~\ref{fig:box-type IC} demonstrates the spatial profile of a box-type initial condition. In particular, this specific initial data can be viewed as two Riemann initial conditions.

Now, to construct an initial condition for the lattice \eqref{e: dmKdV equation} which is consistent with Eq.~\eqref{e: IC for non-regularized and regularized models}, we simply apply the relation in Eq.~\eqref{eq: second set of slow variables} so that
\begin{equation}
    u_n(0) = u_0\left(\epsilon n\right),
\end{equation}
where $u_0$ is defined in Eq.~\eqref{e: IC for non-regularized and regularized models}.

Finally, according to the relation specified in Eq.~\eqref{e: multi-scale ansatz}, the initial condition for the continuum modified KdV equation \eqref{e: continuum mKdV} is given as
\begin{equation}
    u(\xi,0) = \epsilon^{-1}u_n(0),
\end{equation}
where we keep $\epsilon = 0.1$ throughout this work.

\section{Rarefaction waves}\label{sec: RWs}

In this section, we analyze the numerical discrete RWs observed in the simulation of the discrete mKdV equation \eqref{e: dmKdV equation}. In particular, the numerical RWs can be approximated by the self-similar solutions associated with the dispersionless system of the quasi-continuum models. We shall compute these theoretical self-similar solutions for all of the three quasi-continuum models in Eqs.~\eqref{e: continuum mKdV}, \eqref{e: second continuum model} and \eqref{e: regularized model}.

Firstly, for the continuum mKdV equation, we notice its corresponding dispersionless system is given as
\begin{equation}\label{e: dispersionless mKdV}
    u_\tau = 6u^2u_\xi.
\end{equation}
which is equipped with the Riemann initial condition for $u_- > u_+$.

We then look for self-similar solutions in the form of 
\begin{equation}\label{self-similar ansatz for mKdV}
    u(\xi,\tau) = S(\nu), \quad \nu = \xi/\tau.
\end{equation}
Substitution of this ansatz \eqref{self-similar ansatz for mKdV} into Eq.~\eqref{e: continuum mKdV} yields
\begin{equation}\label{self-similar ODE}
    -\nu \partial_{\nu}S = 6S^2\partial_\nu S.
\end{equation}
We then solve for $S$ in Eq.~\eqref{self-similar ODE} to obtain the following analytical self-similar solution
\begin{equation}\label{e: self-similar solution of mKdV}
    u(\xi,\tau) = \begin{cases}
        u_-, \quad \nu < -6u_-^2, \\
        \sqrt{-\frac{\nu}{6}}, \quad -6u_-^2 \leq \nu \leq -6u_+^2,\\
        u_+, \quad \nu > -6u_+^2.
    \end{cases}
\end{equation}
To compare such self-similar solution \eqref{e: self-similar solution of mKdV} with the numerical discrete RW of Eq.~\eqref{e: dmKdV equation}, we transform Eq.~\eqref{e: self-similar solution of mKdV} from the coordinates of $(\xi,\tau)$ specified in Eq.~\eqref{e: multi-scale ansatz} to that of $(n,t)$:
\begin{equation}\label{e: self-similar soln transform back}
   \begin{aligned}
    &u_n(t) = \epsilon u(\xi,\tau) \\
    &= \begin{cases}
        \epsilon u_-, \quad n < -2\left(u_-^2\epsilon^2 + 1\right)t\\
        \sqrt{-\left(\frac{n}{2t} + 1\right)}, \quad -2\left(u_-^2\epsilon^2 + 1\right)t \leq n \leq -2\left(u_+^2\epsilon^2 + 1\right)t,\\
        \epsilon u_+, \quad n > -2\left(u_+^2\epsilon^2 + 1\right)t.
    \end{cases}
    \end{aligned}
\end{equation}

In addition, for the two quasi-continuum models in Eqs.~\eqref{e: second continuum model} and \eqref{e: regularized model}, we note that the two models admit the same dispersionless system which reads
\begin{equation}\label{dispersionless system of two other models}
    u_T = 2\left(1+u^2\right)u_X,
\end{equation}
which is subject to the Riemann initial condition for $u_- > u_+$.

Then, we similarly seek the self-similar solutions to Eq.~\eqref{dispersionless system of two other models} in the following form
\begin{equation}\label{self-similar ansatz for other two models}
    u(X,T) = \tilde{S}(\gamma), \quad \gamma = X/T.
\end{equation}
Substitution of Eq.~\eqref{self-similar ansatz for other two models} into \eqref{dispersionless system of two other models} yields
\begin{equation}\label{self-similar ODE for other two models}
    -\gamma \tilde{S}_\gamma = 2\left(1+\tilde{S}^2\right)\tilde{S}_\gamma.
\end{equation}
Solving Eq.~\eqref{self-similar ansatz for other two models} for $\tilde{{S}}$ yields the following theoretical self-similar solution
\begin{equation}\label{e: self-similar solutions for other two models}
   \begin{aligned}
    &u(X,T) =\\
    &\begin{cases}
        u_-, \quad X < -2\left(1+u_-^2\right)T,\\
        \sqrt{-\left(\frac{X}{2T}+1\right)}, \quad -2(1+u_-^2)T \leq X \leq -2(1+u_+^2)T,\\
        u_+, \quad X > -2(1+u_+^2)T.
    \end{cases}
    \end{aligned}
\end{equation}
Transforming the self-similar solution \eqref{e: self-similar solutions for other two models} from $(X,T)$ \eqref{eq: second set of slow variables} to $(n,t)$ yields
\begin{equation}\label{e: self-similar solution for other two models}
   \begin{aligned}
       &u_n(t) = u(X,T) \\
       &= \begin{cases}
           u_-, \quad n < -2(1+u_-^2)t,\\
           \sqrt{-(\frac{n}{2t}+1)}, \quad -2(1+u_-^2)t \leq n \leq -2(1+u_+^2)t,\\
           u_+, \quad n > -2(1+u_+^2)t.
       \end{cases}
   \end{aligned}
\end{equation}
Interestingly, we observe that the two analytical self-similar solutions in Eqs.~\eqref{e: self-similar solution of mKdV} and \eqref{e: self-similar solution for other two models} coincide with each other.

\begin{figure}[t!]
    \centering
    \includegraphics[width=0.55\linewidth]{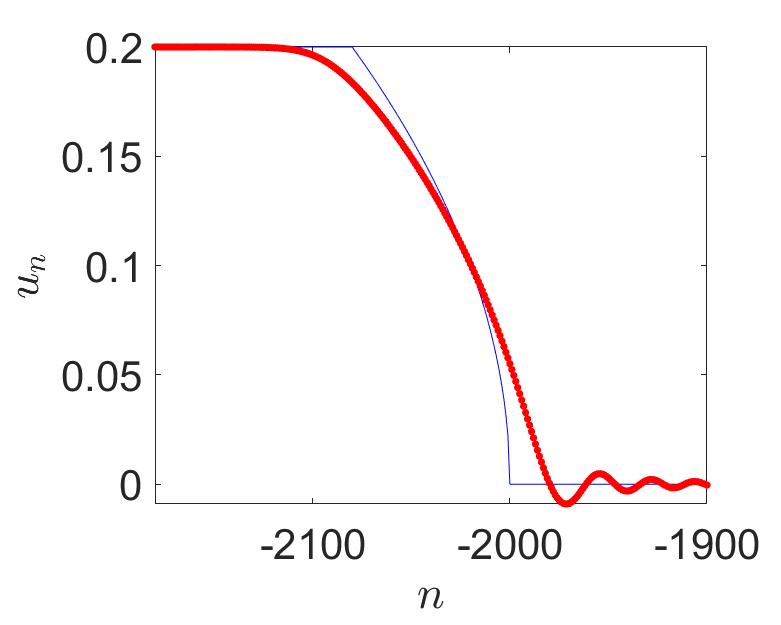}
    \caption{The comparison of the numerical RW (discrete red dots) of the discrete mKdV equation \eqref{e: dmKdV equation} with the self-similar solution (blue curve) of the quasi-continuum models at $t = 1000$ with $u_- = 0.2$ and $u_+ = 0$.}
    \label{fig:RW comparison}
\end{figure}

Fig.~\ref{fig:RW comparison} displays the comparison of the analytical self-similar solution in Eqs.~\eqref{e: self-similar soln transform back} and \eqref{e: self-similar solutions for other two models} with the numerical discrete RW of the lattice \eqref{e: dmKdV equation} at $t = 1000$. From their close alignment, we can see that the quasi-continuum models admit a good performance in approximating the numerical RW.

\section{Numerical validation}\label{sec: numerical vali}

Now, we are ready to discuss the relevant numerical comparisons to examine the performance of these three quasi-continuum models in Eqs.~\eqref{e: continuum mKdV}, \eqref{e: second continuum model} and \eqref{e: regularized model} in approximating the DSWs and their associated edge characteristics. Particularly, we shall compare the DSW-fitting theoretical predictions on DSW edge speeds and amplitude with their associated numerical counterparts. However, before we display the relevant results. We ought to discuss and understand the approaches applied to numerically gauge these edge quantities. Firstly, it is relatively simple to measure the numerical DSW amplitude at time $t$, denoted as $a_-^N(t)$, by computing 
\begin{equation}\label{numerical DSW amplitude}
    a_-^N(t) = \max_{x}\{u(x,t)\} - u_-,
\end{equation}
where $u(x,t)$ is the numerical solution of the relevant models of interest at time $t$. We shall compare the numerical quantity in Eq.~\eqref{numerical DSW amplitude} at the final time snapshot $t_f$ (i.e. $a_-^N(t_f)$) with the DSW-fitting analytical predictions in Eqs.~\eqref{soliton amp-speed rela 1} and \eqref{DSW-fitting prediction on DSW amplitude of continuum mKdV}.

Secondly, regarding the solitonic-edge speed of the DSWs, we first identify the solitonic-edge locations by computing the values of the $x$ coordinate associated with the highest peak of the DSW. Specifically, we shall treat such $x$ coordinates as the location of the solitonic edge of the DSW, denoted as $x_-$. Then, we will obtain a time-series data of the solitonic-edge location as $x_-(t)$ so that the numerical solitonic-edge speed of $s_-^N$ is simply computed as follows:
\begin{equation}\label{numerical solitonic-edge speed}
    s_-^N = \frac{x_-(t_2) - x_-(t_1)}{t_2 - t_1},
\end{equation}
where $t_1 < t_2$ denotes two arbitrary time snapshots.

Next, for the linear-edge speed of $s_+^N$, it can be computed with the same formula specified in Eq.~\eqref{numerical solitonic-edge speed} with the solitonic-edge location of $x_-$ replaced by that of the linear-edge of $x_+$. We shall apply the methods mentioned in \cite{YANG2026103695,Yang_Biondini_Chong_Kevrekidis_2025} to measure the linear-edge locations of the DSW, so we omit the discussion of such details here for brevity.

\begin{figure}[b!]
    \centering
    \includegraphics[width=1.05\linewidth]{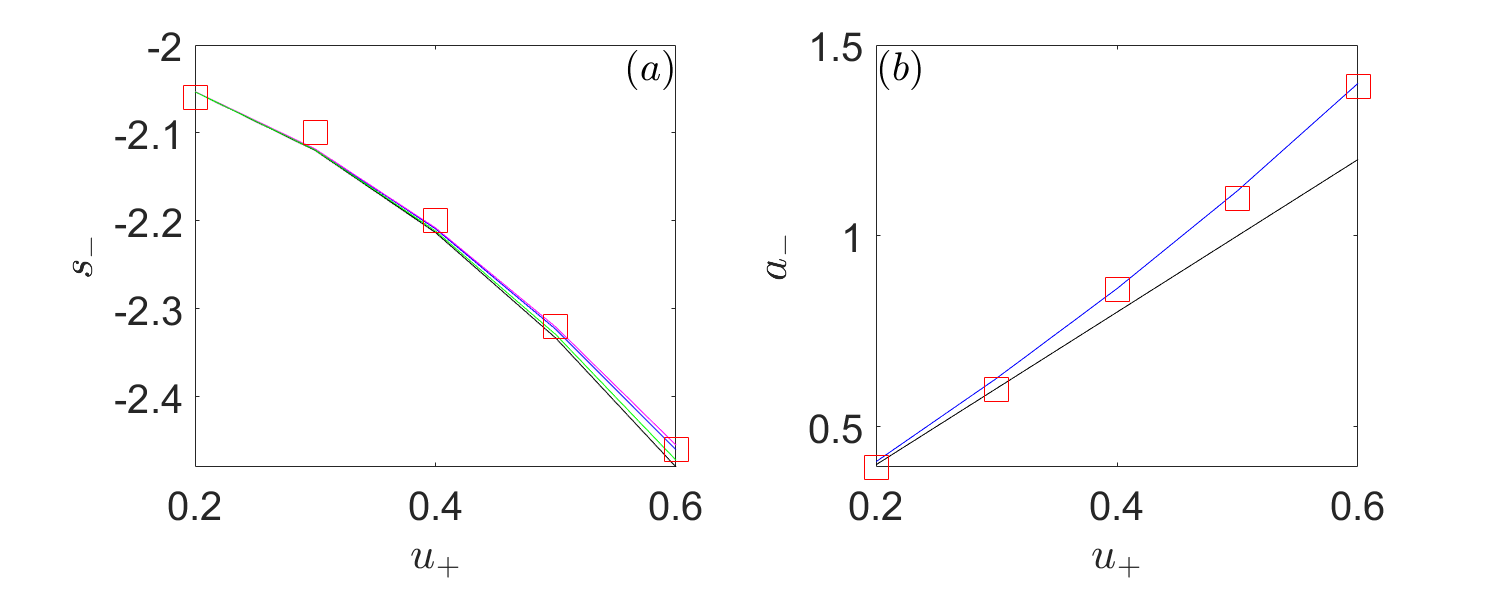}
    \caption{The comparison of the solitonic-edge features including the speed $s_-$ and amplitude $a_-$ of the DSW. Notice that in both panels $(a)$ and $(b)$, we fix $u_- = 0$, while varying $u_+$ from 0.2 to 0.6.}
    \label{fig:solitonic-edge features comparison}
\end{figure}

\begin{figure}[t!]
    \centering
    \includegraphics[width=0.55\linewidth]{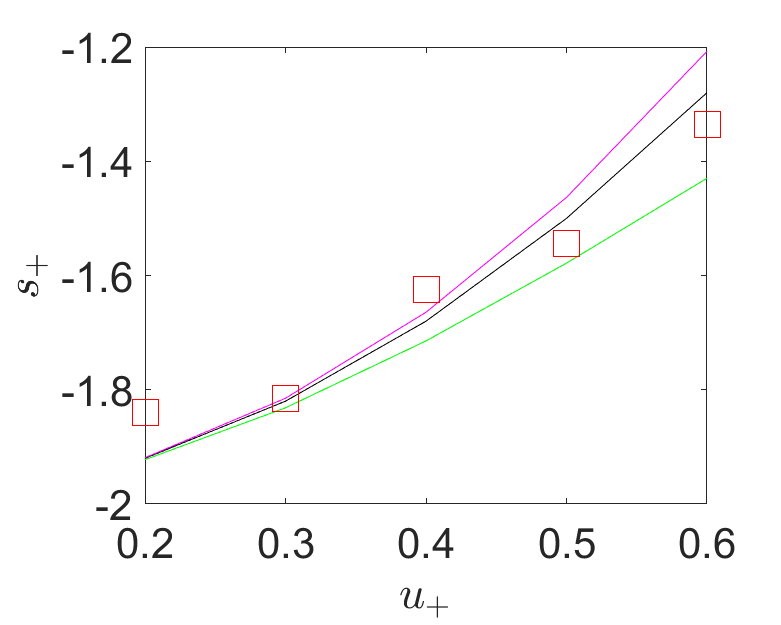}
    \caption{The comparison of the linear-edge speed $s_+$ of the DSW. Notice that we fix $u_- = 0$ and vary the values of $u_+$ from 0.2 to 0.6.}
    \label{fig: comparison of linear-edge speeds}
\end{figure}

\begin{figure}[b!]
    \centering
    \includegraphics[width=0.45\linewidth]{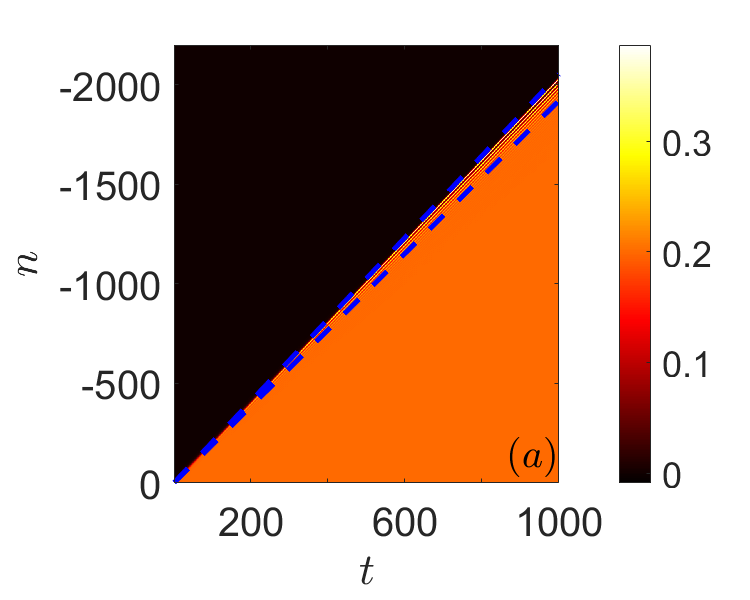}
    \hfill
    \includegraphics[width=0.45\linewidth]{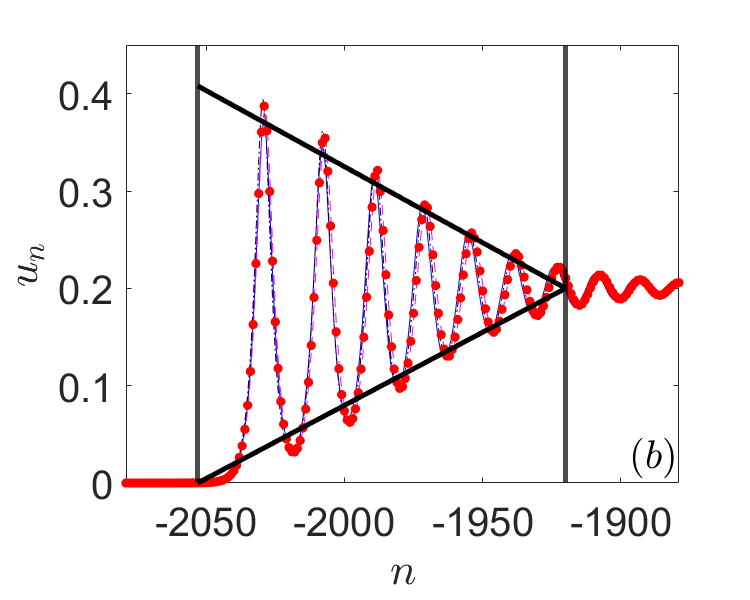}
    \caption{The space-time evolution dynamics $(a)$ and comparison of the spatial profiles $(b)$ of the DSW at $t = 1000$, where $u_- = 0$ and $u_+ = 0.2$.}
    \label{fig:DSW spatial-profile comparison}
\end{figure}

Fig.~\ref{fig:solitonic-edge features comparison} depicts the comparison of the solitonic-edge features of the DSW including its amplitude and speed. We notice that discrete red-square dots refer to the numerically measured solitonic-edge quantities of the discrete DSWs of the lattice \eqref{e: dmKdV equation} via applying the methods mentioned previously. We recall that the overarching goal of the present work is to apply the quasi-continuum models to approximate the discrete DSW from Eq.~\eqref{e: dmKdV equation}, so it suffices to only make numerical estimations on the solitonic-edge quantities of the DSW of the discrete mKdV equation \eqref{e: dmKdV equation} and compare then with the DSW-fitting theoretical predictions from Section \ref{sec: DSW fitting}. Moreover, the solid lines in panels $(a)$ and $(b)$ depict the DSW-fitting predictions associated with the four models in Eqs.~\eqref{e: dmKdV equation} (in blue), \eqref{e: continuum mKdV} (in black), \eqref{e: second continuum model} (in magenta) and \eqref{e: regularized model} (in green). On the one hand, we can clearly see the good agreement of the numerical solitonic-edge speed of the discrete DSW from Eq.~\eqref{e: dmKdV equation} with the DSW-fitting predictions of all four models. On the other hand, for the solitonic-edge amplitude comparison displayed in panel $(b)$ of Fig.~\ref{fig:solitonic-edge features comparison}, we notice that only the two models of Eqs.~\eqref{e: dmKdV equation} and \eqref{e: continuum mKdV} have DSW-fitting predictions due to the explicitly known soliton amplitude-speed relations given in Eqs.~\eqref{soliton amp-speed rela 1} and \eqref{DSW-fitting prediction on DSW amplitude of continuum mKdV}, respectively. Obviously, both DSW-fitting theoretical predictions provide reasonable approximations to the numerically measured amplitude of the discrete DSW of Eq.~\eqref{e: dmKdV equation} provided that the jump of the Riemann initial data is small (i.e. $\Delta \ll 1$), where the jump, denoted as $\Delta$, is defined as $\Delta \equiv \left|u_- - u_+\right|$. When the value of $\Delta$ increases, the theoretical amplitude predicted associated with the continuum mKdV equation \eqref{e: continuum mKdV} starts to deviate from the numerical amplitude. But, the amplitude predicted based on the discrete lattice \eqref{e: dmKdV equation} still agrees very well its numerical counterparts. Next, Fig.~\ref{fig: comparison of linear-edge speeds} shows the comparison of the DSW linear-edge speeds. Such comparison still obeys the pattern that a smaller jump of $\Delta$ leads to better agreement between the theoretical and numerical quantities. Moreover, it is worthwhile to notice that data points of the numerically estimated linear-edge speed of the DSW of Eq.~\eqref{e: dmKdV equation} demonstrate a ``wiggle" feature. This is due to the uncertainty \cite{Yang_Biondini_Chong_Kevrekidis_2025} in measuring the linear-edge locations of the DSW. 

Finally, the Fig.~\ref{fig:DSW spatial-profile comparison} showcases the firstly the space-time evolution dynamics of the discrete DSW of Eq.~\eqref{e: dmKdV equation} in panel $(a)$. In particular, the panel $(a)$ shows a density plot of the discrete DSW of Eq.~\eqref{e: dmKdV equation}, where the dashed blue lines represent the theoretical predictions of the linear and solitonic-edge locations. Namely, they are constructed according to 
\begin{equation}
    n_+(t) = s_+ t, \quad n_-(t) = s_- t,
\end{equation}
respectively, where $s_{\pm}$ are the DSW-fitting predicted edge speeds in Eq.~\eqref{dsw-fitting on lattice}. In addition, the right panel of $(b)$ shows the spatial-profile comparison of the discrete DSW which is represented as discrete red dots with the those of the quasi-continuum models in Eqs.~\eqref{e: continuum mKdV}, \eqref{e: second continuum model} and \eqref{e: regularized model}, which are shown as solid curves. Moreover, we have also drawn a black triangular region according to the DSW-fitting theoretical results, which serves as a theoretical prediction on the spatial profile of the discrete DSW of Eq.~\eqref{e: dmKdV equation}. To construct such region, we first compute the three points: $(n_+, u_+)$, $(n_-,u_-)$ and $(n_-,u_-+a_-)$, where $a_-$ refers to theoretically predicted DSW amplitude based on Eq.~\eqref{soliton amp-speed rela 1}. The upper and lower oblique lines are then constructed by connecting the points of $(n_+, u_+)$ with $(n_-, u_- + a_-)$, and of $(n_+, u_+)$ with $(n_-, u_-)$. The two vertical lines of the triangular region at the linear and solitonic edges of the DSW are $n = n_+$ and $n = n_-$, respectively. As a result, the close alignment of the spatial profile of the discrete DSW of the discrete mKdV equation \eqref{e: dmKdV equation} with those of all the quasi-continuum models demonstrate the good performance of each quasi-continuum model in approximating the discrete DSW.

\section{Conclusions and future directions}\label{sec: conclusions}

In this paper, we revisit the discrete modified KdV equation and derive three quasi-continuum models which are the continuum modified KdV equation \eqref{e: continuum mKdV}, non-regularized model \eqref{e: second continuum model} and regularized model \eqref{e: regularized model}. We also review the analytical periodic traveling waves of the discrete modified KdV equation \eqref{e: dmKdV equation}. We then derive the Whitham modulation equations associated with the two quasi-continuum models in Eqs.~\eqref{e: second continuum model} and \eqref{e: regularized model} and perform reductions on both modulation system to arrive at the DSW-fitting simple-wave ODEs. We then solve the system of simple-wave ODEs euippbed with appropriate initial conditions to obtain the theoretical predictions on different edge features of the DSWs of the quasi-continuum models. Numerical comparisons are conducted, demonstrating the good performance of the approximations made by all the quasi-continuum models. 

However, there are many interesting open questions related to the present work, and we here only list a few of them. Firstly, we notice that since the discrete modified KdV equation is shown to be integrable \cite{CHEN2023133652}, then it is possible to perform Whitham analysis directly on Eq.~\eqref{e: dmKdV equation} itself. Namely, in principle, one can find the associated conservation laws of Eq.~\eqref{e: dmKdV equation} to perform Whitham averaging in order to obtain the corresponding Whitham modulation equations. Moreover, whenever a model is integrable, it is also possible to diagonalize the Whitham modulation system by casting it into the equivalent Riemann-invariant form. One can thereafter utilize the diagonalized Whitham modulation system to construct the theoretical solution for the discrete DSW of Eq.~\eqref{e: dmKdV equation} so that this shall provide an exact analytical description on the discrete DSW. Secondly, it is also interesting to investigate the well-posedness of the two proposed quasi-continuum models in Eqs.~\eqref{e: second continuum model} and \eqref{e: regularized model} via analytical approaches. These new directions are currently under consideration and will be reported in future publications.

\bibliography{main}

\bibliographystyle{unsrt}

\end{document}